\documentclass[aps,prd,twocolumn,preprintnumbers,showpacs,nofootinbib,amssymb]{revtex4}
\usepackage{graphicx}
\usepackage{amsmath}
\usepackage{amssymb}
\usepackage{amsfonts}
\usepackage{bm}
\usepackage{color}

\def\be{\begin{equation}}
\def\ee{\end{equation}}
\def\bea{\begin{eqnarray}}
\def\eea{\end{eqnarray}}

\begin{document}

\title{Gravitational phase transitions and instabilities of
self-gravitating fermions\\
 in general relativity}
\author{Pierre-Henri Chavanis}
\affiliation{Laboratoire de Physique Th\'eorique, Universit\'e de Toulouse,
CNRS, UPS, France}
\author{Giuseppe Alberti}
\affiliation{Laboratoire de Physique Th\'eorique, Universit\'e de Toulouse,
CNRS, UPS, France}
\affiliation{
Living Systems Research, Roseggerstra\ss e 27/2, A-9020 Klagenfurt am
W\"{o}rthersee,
Austria}

\begin{abstract}

We discuss the occurrence of gravitational phase
transitions and instabilities in a gas of 
self-gravitating fermions within the framework of general relativity. In the
classical (nondegenerate) limit, the system undergoes a gravitational collapse
at low
energies $E<E_c$ and low temperatures $T<T_c$. This is called ``gravothermal
catastrophe'' in the microcanonical ensemble and ``isothermal collapse'' in
the canonical ensemble. When quantum mechanics is taken into account and
when the particle number is below the Oppenheimer-Volkoff limit ($N<N_{\rm
OV}$), complete
gravitational collapse is prevented by the Pauli exclusion principle. In that
case, the Fermi gas undergoes a gravitational phase transition from a gaseous
phase to a condensed phase. The condensed phase represents a compact object
like a white dwarf,
a neutron star, or a dark matter fermion ball. When $N>N_{\rm
OV}$, there can be a
subsequent gravitational collapse
below a lower critical energy $E<E''_c$ or a lower critical temperature
$T<T'_c$  leading
presumably  to the formation of a
black hole. The evolution of the system is different in the microcanonical  and
canonical  ensembles. In the microcanonical ensemble, the system takes a
``core-halo'' structure. The  core consists in a 
compact quantum object or a
black hole while the hot halo is expelled at large
distances. This is reminiscent of the red giant structure of low-mass stars or
the implosion-explosion of massive stars (supernova). In the canonical ensemble,
the
system collapses as a whole
towards a compact object or a black hole. This is reminiscent of the implosion
of supermassive stars (hypernova).

\end{abstract}


\maketitle

\section{Introduction}

The study of phase transitions is an important problem in physics. Some examples
include solid-liquid-gas phase transitions,  superconducting and superfluid
transitions, Bose-Einstein condensation, liquid-glass phase transition in
polymers, liquid crystal phases, Kosterlitz-Thouless transition etc.
Self-gravitating systems also undergo phase transitions but they are special
because of the unshielded long-range attractive nature of the gravitational
interaction \cite{paddy,katzrevue,ijmpb}. This leads to unusual phenomena such
as quasistationary states, negative
specific heats, ensembles inequivalence,  long-lived
metastable states, and
gravitational collapse. The statistical mechanics of systems with long-range
interactions is also an important topic in statistical mechanics
\cite{houches,cdr,campabook} because of its peculiarities and its various 
applications
in astrophysics,
plasma physics and hydrodynamics.

In this Letter, we consider the statistical mechanics of self-gravitating
fermions in general relativity. This is a fundamental problem that combines
general relativity, quantum mechanics and statistical mechanics. In addition,
the
thermodynamics of the self-gravitating
Fermi gas can
have application in relation to the physics of white dwarfs, neutron stars and
dark matter halos made of massive neutrinos. It may also be related, as we shall
see, to the supernova and hypernova phenomena. It represents therefore a topic
of considerable interest both from a fundamental and an astrophysical point of
view.

The statistical mechanics of self-gravitating systems dates back to
the works of Antonov \cite{antonov} and Lynden-Bell and Wood \cite{lbw} who
considered nonrelativistic classical stellar systems such as globular clusters.
They
enclosed the gas of stars within a spherical box of radius $R$ in order to
prevent its
evaporation. They
showed the absence of
a statistical equilibrium state below a critical energy $E_c=-0.335GM^2/R$ in
the
microcanonical ensemble (MCE) or below a critical temperature $T_c=GMm/2.52k_B
R$ in
the canonical
ensemble (CE) (the  critical temperature $T_c$ was found earlier by Emden
\cite{emden}). The
gravitational
collapse in the MCE is called ``gravothermal
catastrophe'' \cite{lbw}. In the case of globular clusters, it leads ultimately 
to the formation of a binary star surrounded by a hot halo \cite{inagakilb}
(this structure has
an infinite entropy $S\rightarrow +\infty$ at fixed energy \cite{sc}).
The gravitational
collapse in the CE
is called ``isothermal collapse'' \cite{aaiso}. In the case of self-gravitating
Brownian particles it leads to the formation of a Dirac peak containing all the
mass \cite{post} (this structure has an infinite free energy $F\rightarrow
-\infty$ \cite{sc}).

We may expect that quantum mechanics will prevent complete gravitational
collapse on account of the Pauli exclusion principle (for fermions) giving rise
to an additional quantum pressure. This is indeed the
case for a nonrelativistic system of fermions as shown by Fowler \cite{fowler},
Stoner \cite{stoner29} and Chandrasekhar \cite{chandra31nr} at $T=0$ in the
context of white dwarf stars. The statistical mechanics
of self-gravitating fermions at finite temperature in the nonrelativistic limit
was developed by
Hertel and
Thirring \cite{ht}, Bilic and Viollier \cite{bvn} and Chavanis \cite{pt,ijmpb}.
They
showed that a gas of fermions experiences a gravitational phase transition from
a gaseous phase to a
condensed phase when the energy or the temperature passes below a critical
value. In that case, an equilibrium state (gaseous or condensed) exists for all
accessible values of temperature and energy. However,
the situation is expected to change for the general relativistic Fermi gas 
because we already know from the work of Oppenheimer and Volkoff \cite{ov}
on neutron stars that,
at $T=0$,
there is no equilibrium state if the particle number $N$ overcomes
the Oppenheimer-Volkoff limit $N_{\rm
OV}=0.39853\, \left ({\hbar c}/{G}\right
)^{3/2}m^{-3}$.\footnote{This is the general relativistic extension of the
Anderson-Stoner-Chandrasekhar-Landau
\cite{anderson,stoner30,chandra31,landau32} 
maximum mass of special relativistic Newtonian white dwarf stars.}

In this Letter, we report the results of our recent investigations
concerning
the statistical mechanics of self-gravitating fermions at finite temperature in
the framework of general relativity. They complete the previous works of Bilic
and Viollier \cite{bvrelat,bvr} and Roupas \cite{roupas1,roupas,roupasprd}. The
complete
study being extremely rich,
all the details are given in a series of exhaustive papers
\cite{acb,acf,rgf,rgb}. These
papers
also contain a detailed historic of the subject with a long list of
references. In this Letter, we emphasize the main results of our study and refer
to our companion papers for more details and additional results.

\section{Basic equations}

The statistical equilibrium state of a self-gravitating system in
general relativity is obtained by maximizing the entropy 
$S=-k_B\int
C(f) e^{\lambda(r)/2} 4\pi r^2\, d{r}d{\bf p}$ at fixed
mass-energy 
$Mc^2=\int f E(p) 4\pi r^2\, drd{\bf p}$ and particle number
$N=\int f e^{\lambda(r)/2} 4\pi r^2\, d{r}d{\bf p}$ \cite{bvrelat,roupas1,rgf}.
Here $f({\bf r},{\bf p})$ is the distribution function (DF) and
$e^{\lambda(r)/2}
4\pi r^2\, d{r}$ is
the proper volume element involving the metric coefficient
$e^{\lambda(r)/2}=[1-2GM(r)/rc^2]^{-1/2}$ where $M(r)c^2$ is the mass-energy
contained within the sphere of radius $r$. For fermions, $S$ is the Fermi-Dirac
entropy with
$C(f)=f\ln(f/f_m)+(f_m-f)\ln(1-f/f_m)$ with $f_m=2/h^3$. In the
nondegenerate (classical) limit $f\ll f_m$, the Fermi-Dirac entropy
reduces to
the Boltzmann entropy with
$C(f)=f[\ln(f/f_m)-1]$. The extremization problem leads to 
the Fermi-Dirac DF
\begin{equation}
\label{e1}
f({\bf r},{\bf p})=\frac{2}{h^3}\frac{1}{1+e^{-\alpha}e^{E(p)/k_B T(r)}},
\end{equation}
where
$E(p)=\sqrt{p^2c^2+m^2c^4}$ is the energy
of a
particle
and $T(r)$ is the local temperature. In general relativity, the temperature is
spatially inhomogeneous even at statistical equilibrium (Tolman's effect)
\cite{tolman}. The extremization of entropy at fixed mass-energy and particle
number also yields the
Tolman-Oppenheimer-Volkoff (TOV) equations \cite{tolman,ov}
\begin{equation}
\label{e3b}
\frac{dM}{dr}=\frac{\epsilon}{c^2} 4\pi r^2,
\end{equation}
\begin{equation}
\label{e3}
\frac{1}{T}\frac{{\rm
d}T}{{\rm
d}r}=\frac{1}{\epsilon+P}\frac{dP}{dr}=-\frac{1}{c^2}\frac{\frac{GM(r)}{r^2}
+\frac { 4\pi
G}{c^2}Pr}{1-\frac{2GM(r)}{r c^2}},
\end{equation}
expressing the condition of hydrostatic equilibrium,  and the Tolman-Klein
\cite{tolman,klein}
relations $T(r)=T_{\infty} e^{-\nu(r)/2}$ and  $\mu(r)=\mu_{\infty}
e^{-\nu(r)/2}$ relating the local temperature $T(r)$ and the
local chemical potential $\mu(r)$ to the metric coefficient $\nu(r)$
($T_{\infty}$ and $\mu_{\infty}$ are the temperature and the chemical
potential measured by an observer at infinity).
Their ratio $\alpha=\mu(r)/k_B T(r)=\mu_{\infty}/k_B T_{\infty}$ is constant
since they are redshifted
in the same manner. From the DF (\ref{e1}), we can express the particle number
density $n(r)=\int
f\, d{\bf p}$, the energy density $\epsilon(r)=\int f E\, d{\bf p}$ and the
pressure $P(r)=(1/3)\int f p E'(p)\, d{\bf p}$ in terms of $T(r)$ and $\alpha$.
We can then obtain $T(r)$ by solving the TOV equations (\ref{e3b}) and
(\ref{e3}) with the boundary conditions $T(0)=T_0\ge 0$ and $M(0)=0$, and
determining $\alpha$ by the constraint $N(R)=N$. The
mass and the Tolman (global)
temperature are then given by $M=M(R)$ and 
$T_{\infty}=T(R)\sqrt{1-2GM/Rc^2}$. Finally, we can plot  the
caloric curve $T_{\infty}(Mc^2)$ for a given value of $N$ by varying $T_0$ from
$0$ to $+\infty$. Instead of the mass-energy $Mc^2$ it is better to use the
binding energy
$E=(M-Nm)c^2$ which reduces to the Newtonian energy $E=E_{\rm
kin}+W$ in the limit
$c\rightarrow +\infty$. Finally, we introduce the dimensionless energy 
$\Lambda=-ER/GN^2m^2$ and the dimensionless inverse temperature
$\eta=\beta_{\infty}GNm^2/R$ with $\beta_{\infty}=1/k_B T_{\infty}$.

\section{Classical systems}

\begin{figure}
\begin{center}
\includegraphics[clip,scale=0.3]{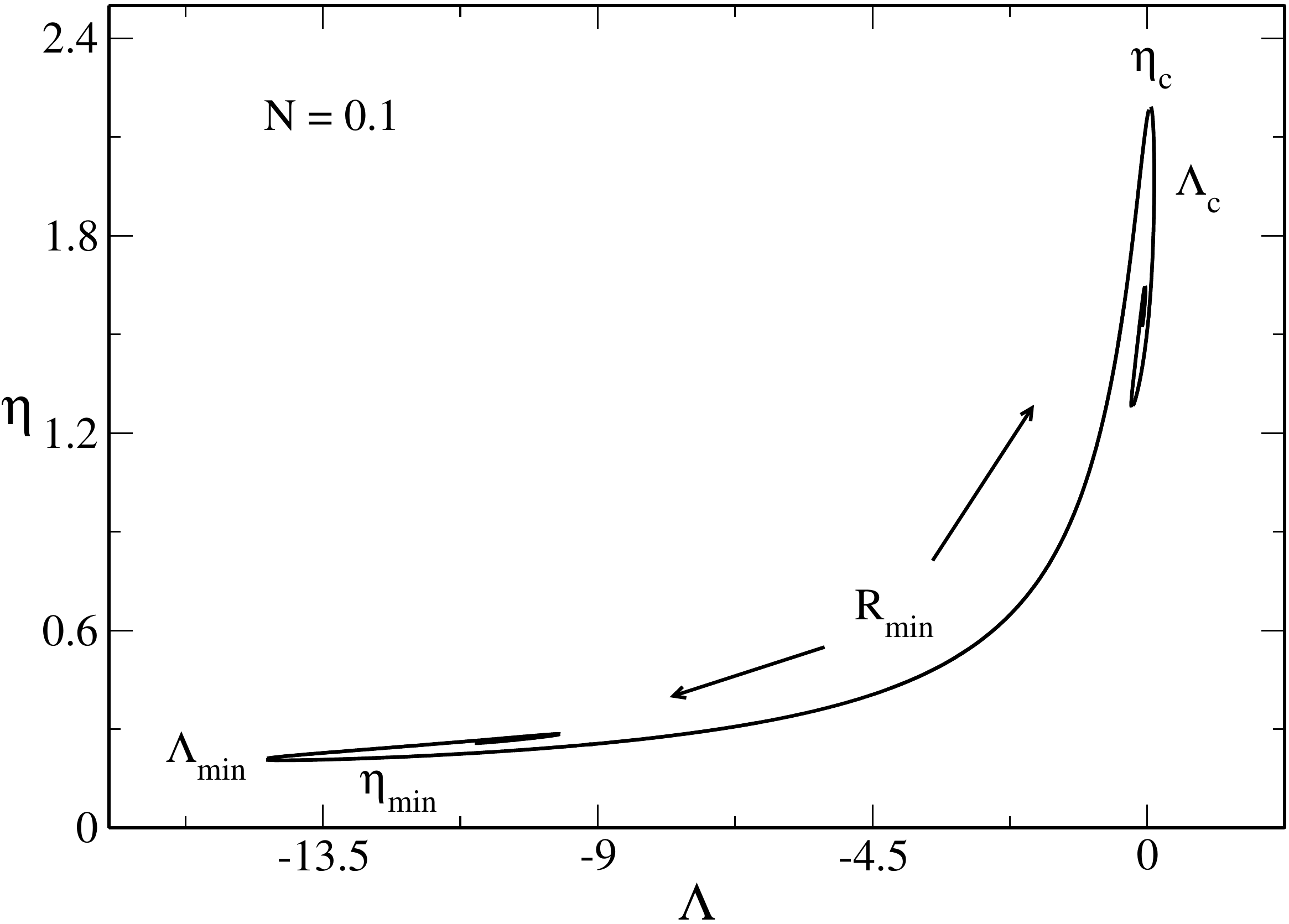}
\caption{Caloric
curve of the general relativistic classical self-gravitating
gas for $\nu=0.1$. It presents a double spiral. The
system collapses at
low energies and low temperatures as in the case of a nonrelativistic classical
self-gravitating gas in Newtonian gravity (cold spiral). It also collapses
(towards a
black hole) at high
energies and high temperatures as in the case of the
self-gravitating black-body radiation (hot spiral).}
\label{kcal_N01_linked_colorsPH}
\end{center}
\end{figure}

The caloric curve $\eta(\Lambda)$ of a classical self-gravitating gas in general
relativity
depends on a single parameter $\nu=GNm/Rc^2$ called the compactness parameter.
It is plotted in Fig.
\ref{kcal_N01_linked_colorsPH} for $\nu=0.1$. It has the
form of a double
spiral parametrized by the energy density contrast ${\cal
R}=\epsilon(0)/\epsilon(R)$ \cite{roupas,acb}. The density contrast ${\cal R}$
is minimum at the
``center'' of the caloric curve and increases along the series of equilibria in
the directions of the spirals.

The cold spiral (on the right) is a relativistic generalization of the caloric
curve obtained by Katz
\cite{katzpoincare1} for the nonrelativistic classical
self-gravitating gas in
Newtonian
gravity. It corresponds to weakly  relativistic
configurations (except when
$\nu$ is large).  It exhibits a minimum energy in the MCE at
$E_c$ and a minimum temperature in the  CE at $T_c$ below which the system
undergoes a gravitational collapse as in the nonrelativistic case.
The hot spiral is a purely general
relativistic result.
It is similar (but not identical \cite{rgb}) to the caloric curve
obtained by Chavanis \cite{aarelat2} for the self-gravitating black-body
radiation. It
corresponds to strongly relativistic
configurations.  It exhibits a
maximum energy in the MCE at $E_{\rm max}$ and a maximum temperature in the  CE
at
$T_{\rm max}$ above which the system undergoes a gravitational collapse leading
presumably to the formation of a black hole.\footnote{Below
$T_c$ the system collapses because it is too cold and the thermal pressure
cannot balance the gravitational attraction. By contrast, above $E_{\rm max}$
the system
collapses because it is too hot and feels ``the weight of heat''
\cite{tolman} (energy is mass so that it gravitates).}

In the CE the series of equilibria is stable on the main branch between
$\eta_{\rm
min}$ and
$\eta_{c}$. According to the Poincar\'e criterion \cite{poincare,katzpoincare1},
it becomes unstable
at the first 
turning points of temperature $\eta_{\rm
min}$ and
$\eta_{c}$ where the specific heat is infinite passing
from positive to negative values. A new  mode of instability is lost at each
subsequent turning
point of
temperature as the spirals rotate clockwise. In the MCE the series of equilibria
is
stable on the main branch between $\Lambda_{\rm min}$ and
$\Lambda_{c}$. According to the Poincar\'e criterion
\cite{poincare,katzpoincare1}, it becomes
unstable at the first
turning points of energy $\Lambda_{\rm min}$ and
$\Lambda_{c}$ where the specific heat vanishes passing from
negative to
positive values.\footnote{Stable equilibrium states may have a negative
specific heat in the MCE (isolated systems with fixed energy) while
this is not possible in the CE (systems in contact with a heat bath with fixed
temperature).} A new  mode of instability is lost at each subsequent turning
point of
energy as the spirals rotate clockwise. There are two regions of ensembles
inequivalence, one on each spiral, between the turning points of
temperature and energy,
i.e., in the first region of negative specific heat.

It has to be noted that the stable equilibrium states are in fact
{\it metastable}
as there is no global maximum of entropy at fixed energy or global minimum of
free energy for
classical self-gravitating systems \cite{antonov,lbw}. However, these metastable
states have a tremendously long lifetime, scaling as $e^N$, so they are
stable in
practice \cite{metastable}.\footnote{Only a large random
fluctuation can drive the system out of a local maximum
of entropy (or local minimum of free energy). This is a rare event. The time
scale for this
phenomenon is exponentially large.}

The
evolution of the caloric curve with $\nu$ is
described in detail in \cite{roupas,acb}. As $\nu$ increases, the cold and hot
spirals
approach each other, merge at $\nu'_S=0.128$, form a loop above
$\nu_S=0.1415$, reduce to a point at $\nu_{\rm max}=0.1764$,
and finally
disappear. The limit $\nu\rightarrow 0$ depends on the normalization that we use
to plot the caloric curve. If we use the normalized variables
$\Lambda=-ER/GN^2m^2$ and $\eta=\beta_{\infty}GNm^2/R$ appropriate to the
nonrelativistic limit we find that
the hot spiral is rejected at infinity and we obtain a limit curve
corresponding to the caloric curve of the nonrelativistic classical
self-gravitating gas. It has the form of a (cold) spiral with turning points at
$\Lambda_c=0.335$ and
$\eta_c=2.52$ \cite{antonov,lbw,katzpoincare1}. Alternatively,  if we use the
normalized variables ${\cal M}=GM/Rc^2$ and ${\cal
B}=\beta_{\infty}Rc^4/GN$  appropriate to the
ultrarelativistic limit  we
find that the cold spiral is rejected at infinity and we obtain a limit
curve corresponding to the caloric curve of the ultrarelativistic classical
self-gravitating gas \cite{rgb}. It has the form of a (hot) 
spiral with turning points at ${\cal M}_{\rm
max}=0.24632$ (like for the self-gravitating black-body radiation
\cite{sorkin,aarelat2})
and
${\cal
B}_{\rm min}=17.809$ (different from the self-gravitating black-body
radiation \cite{rgb}).

{\it Remark--} Let us assume here that the previous results apply to star
clusters (they may also apply to gaseous stars as discussed in Sec. \ref{sec_aa}
in which
case the following discussion would be different).
In Newtonian gravity, it can be shown that all the isotropic DFs of the
form $f=f(\epsilon)$ with $f'(\epsilon)<0$, where $\epsilon=v^2/2+\Phi(r)$ is
the energy of a star by unit of mass, are dynamically stable with respect to the
Vlasov-Poisson equations (which describe collisionless stellar systems), even
those that are thermodynamically unstable
\cite{doremus71}. On the other hand, in general relativity, Ipser \cite{ipser80}
has shown that
dynamical stability with respect to the Vlasov-Einstein equations coincides with
thermodynamical stability so that the equilibrium states located after the
turning
points of energy $E_c$ and $E_{\rm max}$ are both thermodynamically and
dynamically unstable. To solve this apparent paradox, it is
expected that the growth rate of the dynamical instability decreases as the
relativity level decreases. From these considerations, one expects
that
the collapse at the turning point $E_c$ of the cold spiral (weakly relativistic
configurations)
is mainly a
thermodynamical -- collisional -- instability (gravothermal catastrophe) that
takes place on a long, secular, timescale while
the
collapse at the turning point $E_{\rm max}$ of the hot spiral  (strongly
relativistic configurations) is mainly a dynamical -- collisionless -- 
instability that takes place
on a short, dynamical,
timescale.

\section{Quantum systems}

The caloric curve $T_{\infty}(E)$ of a quantum gas of fermions in
general relativity depends on two parameters, the box radius $R$ and the
particle number $N$ (throughout
the paper, the radius is measured in terms of
$R_*=(\hbar^3/m^4Gc)^{1/2}=0.11447\, R_{\rm OV}$ and the
particle number  is measured in terms of
$N_*=(\hbar^3c^3/m^6G^3)^{1/2}=2.5092\, N_{\rm OV}$).

\subsection{$N$-shape structure}

We first consider a radius in the range $R_{\rm CCP}=12.0<R<R_{\rm MCP}=92.0$
so that the caloric curve may display canonical, but not microcanonical,
phase transitions \cite{acf}.

For $N\ll N_{\rm OV}=0.39853$ we are in the nonrelativistic
limit considered in \cite{ht,bvn,pt,ijmpb}. When $N<N_{\rm CCP}(R)$, where 
$N_{\rm CCP}(R)\simeq 2125/R^3$ corresponds
to a canonical
critical point,  the
caloric curve is monotonic like in Fig. 14 of
\cite{ijmpb} for $\mu=10$ and there is no phase transition. When 
$N_{\rm CCP}(R)<N<N_{\rm OV}$ the caloric curve has an $N$-shape
structure similar to Fig. 23 of \cite{ijmpb}. In the MCE,
the whole
series of equilibria is stable. It displays  a region of negative
specific heat $C=dE/dT_{\infty}<0$ between $T_c$ and $T_*$. In the CE, the
region of
negative specific heat
is replaced by a phase transition from the gaseous phase to the condensed
phase. The condensed phase corresponds to a fermion ball (similar to a white
dwarf or a neutron star) with only a tiny atmosphere. In principle, we expect
a first order phase transition to take place at the transition temperature $T_t$
at which the two phases have the same free energy.\footnote{The
canonical transition temperature $T_t$ can also be determined by a method
analogous
to that of Maxwell's plateau in the discussion of the van der Waals gas.} It
would be marked by
a discontinuity of energy. However, this first order
phase transition does not
take place in practice because the metastable gaseous states with $T_c<T<T_t$
(which are local but not global minima of free energy) have tremendously long
lifetimes,
scaling as $e^N$ \cite{metastable}. Not only they are stable in practice, but
they
are also more
physical than strictly stable equilibrium states (global minima of free energy)
which can be reached only if the system spontaneously crosses a huge barrier of
free energy by forming a ``critical droplet'' represented here by a fermion
ball (quantum core), a
very unlikely process (rare event) in the CE. As a result, the physical phase
transition does not take place at $T_t$ but rather at $T_c$ (spinodal point),
the temperature
at which the metastable gaseous branch disappears. At that point, the system
undergoes a gravitational collapse (isothermal collapse) and becomes denser and
denser until quantum effects come into play and
stabilize the system. Once in
the condensed phase, if the temperature increases, there is an explosion,
reverse to the collapse,
from the condensed
phase to the gaseous phase at $T_*$ (spinodal point). We can in this manner
generate an hysteresis cycle in the CE \cite{ijmpb}. 

\begin{figure}
\begin{center}
\includegraphics[clip,scale=0.3]
{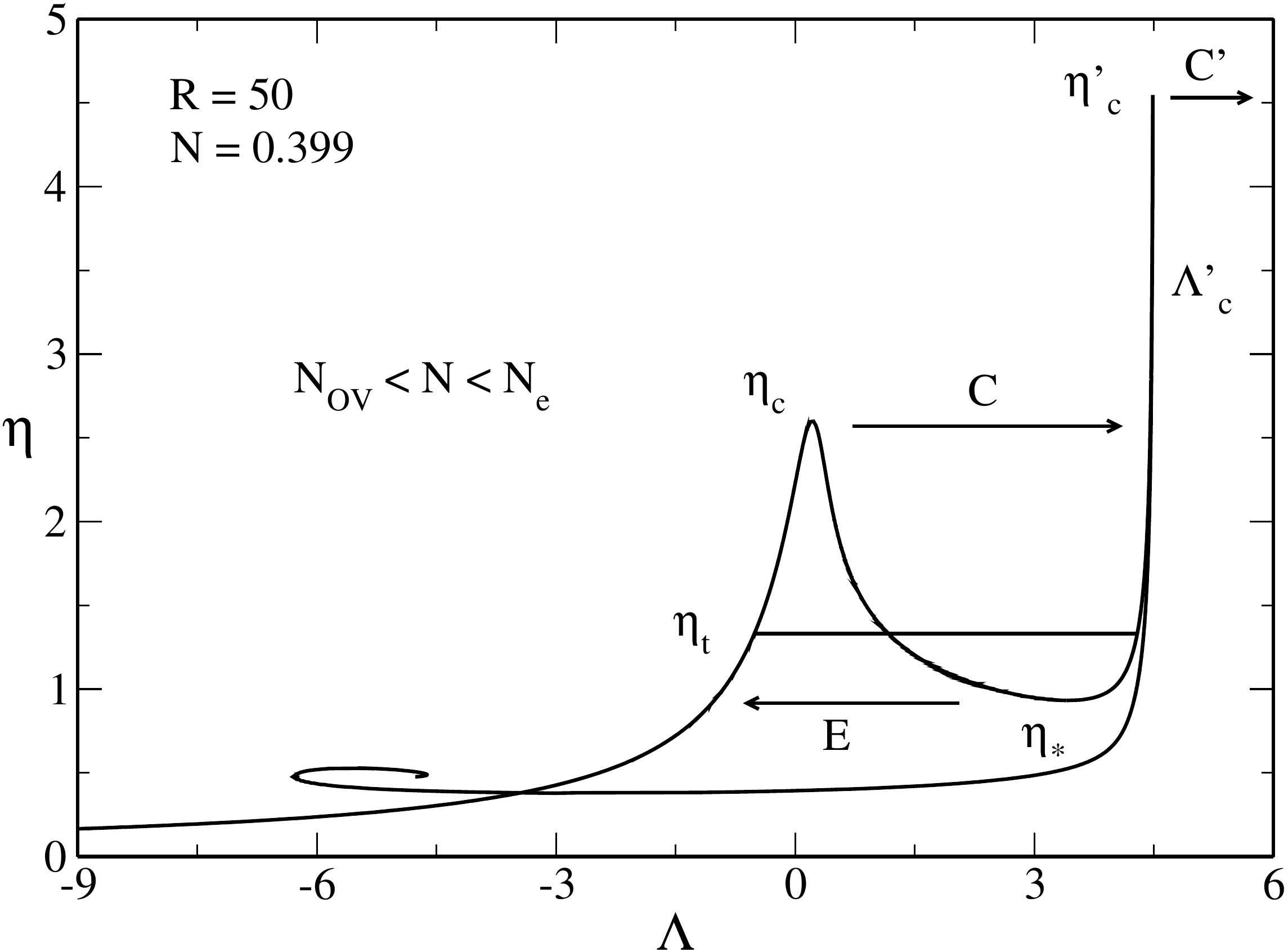}
\caption{Caloric curve for  $N_{\rm OV}<N\ll N_{\rm max}$
(specifically $R = 50$ and $N = 0.399$). The arrows
refer to the CE: (C) collapse at $T_c$ from the gaseous phase to
the condensed phase (fermion ball); (C') collapse at $T'_c$ from the
condensed phase (fermion ball) to a black hole; (E) explosion at $T_*$ from the
condensed phase (fermion ball)  to the gaseous phase.}
\label{kcal_R50_N0p399_unifiedPH}
\end{center}
\end{figure}

\begin{figure}
\begin{center}
\includegraphics[clip,scale=0.3]
{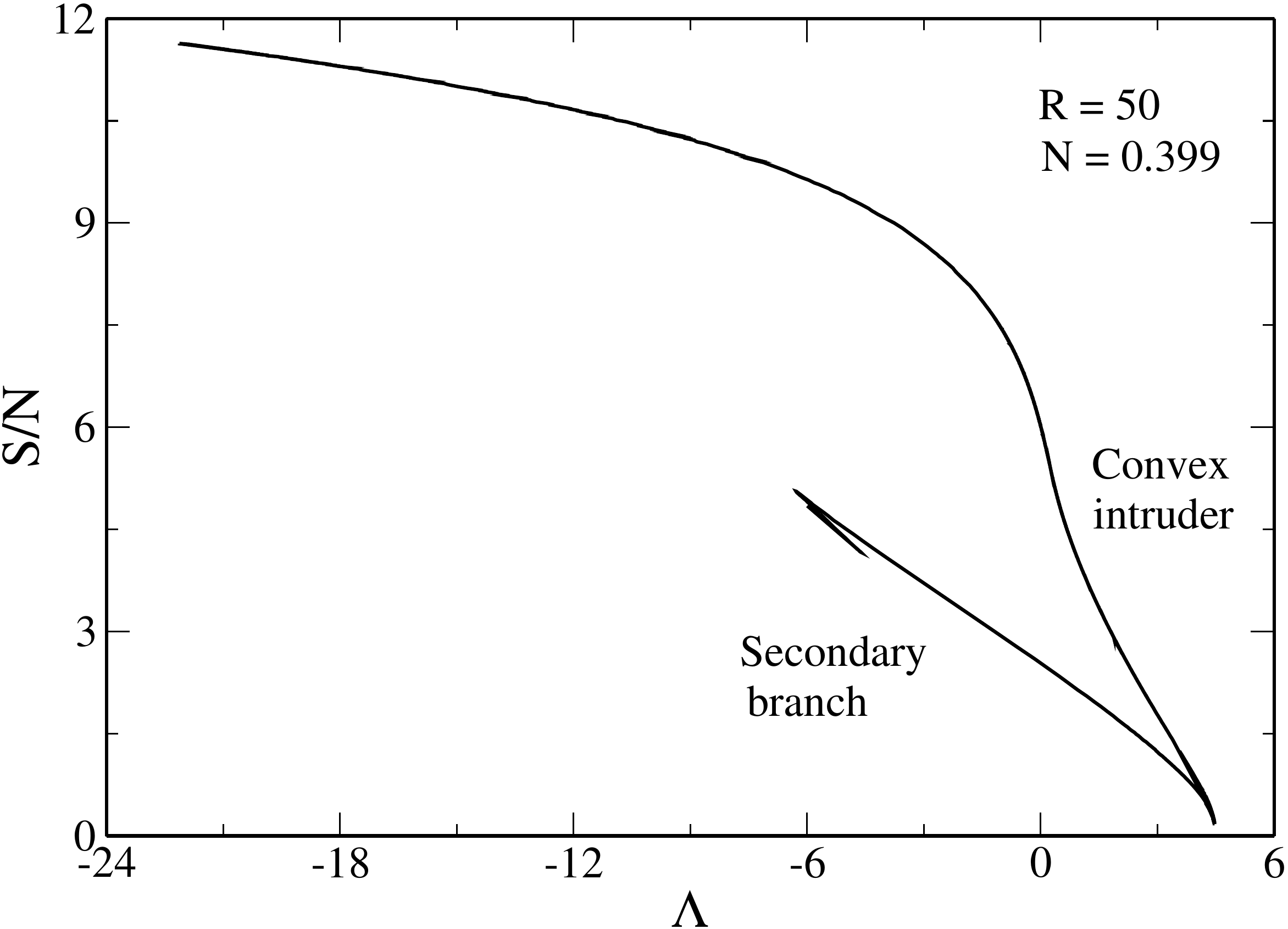}
\caption{Entropy per fermion as a function of the normalized energy for
$N_{\rm OV}<N\ll N_{\rm max}$ (specifically $R = 50$
and $N = 0.399$).}
\label{SLambda_R50_N0p399_colorPH}
\end{center}
\end{figure}

\begin{figure}
\begin{center}
\includegraphics[clip,scale=0.3]{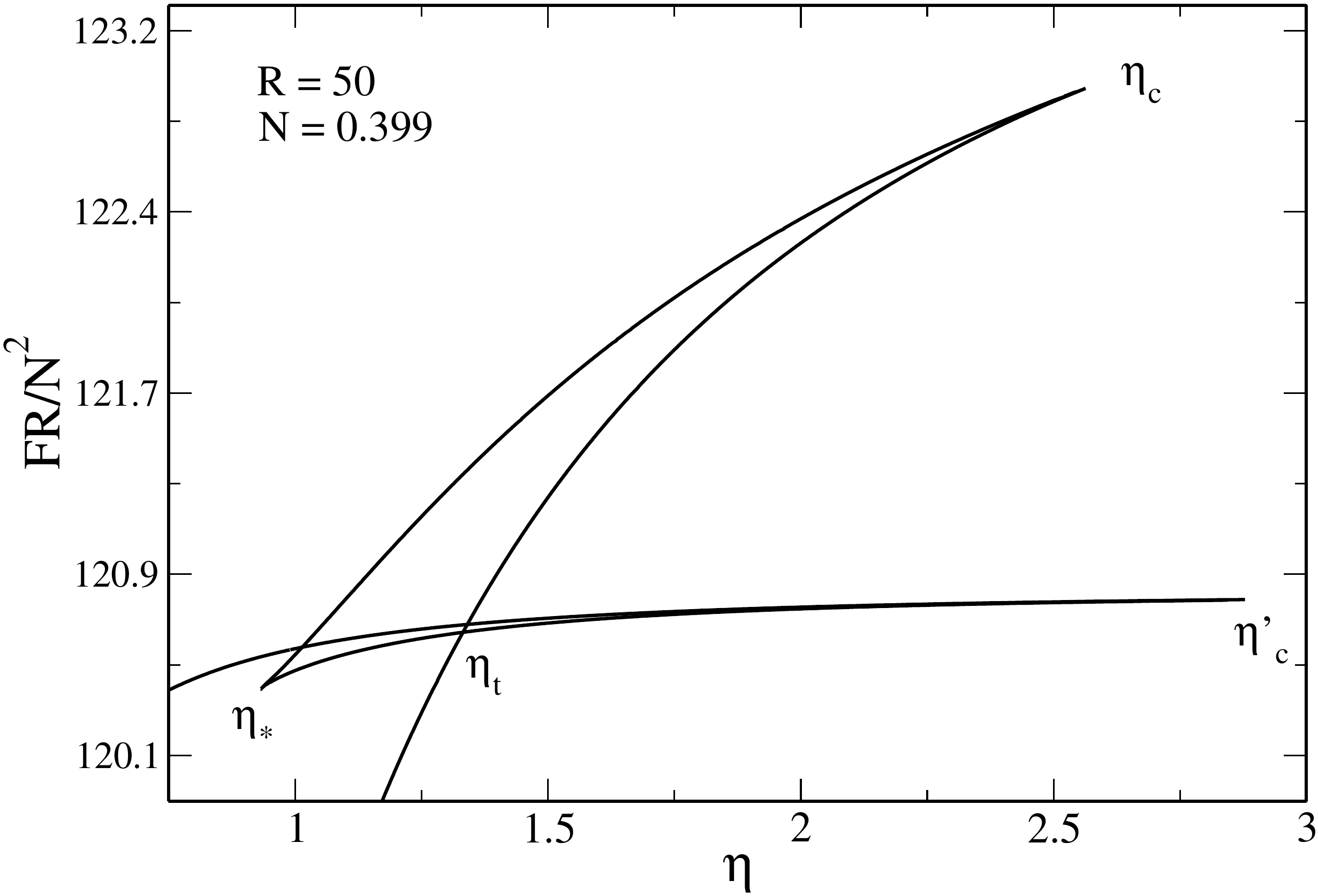}
\caption{Normalized free energy as a function of the normalized inverse
temperature for
$N_{\rm OV}<N\ll N_{\rm max}$ (specifically $R = 50$
and $N = 0.399$).}
\label{Feta_R50_N0p399PH}
\end{center}
\end{figure}

When $N<N_{\rm OV}$,
there is an
equilibrium state (gaseous or condensed) for any accessible value of energy
$E\ge E_{\rm min}$ or temperature $T\ge 0$. Quantum mechanics (Pauli's
exclusion principle) prevents gravitational
collapse of classical self-gravitating systems below $E_c$ or below $T_c$.
However, as $N$ approaches $N_{\rm OV}$, general
relativistic
effects
come into play. For $N_{\rm OV}<N\ll N_{\rm max}$, the caloric curve
has the form of Fig. \ref{kcal_R50_N0p399_unifiedPH}. In the MCE,
there is no equilibrium state below a minimum energy $E'_{c}$. In
that case the system is expected to collapse towards a black hole. The series of
equilibria is stable until $E'_{c}$ and becomes unstable afterwards. The
entropy versus energy curve is plotted in
Fig. \ref{SLambda_R50_N0p399_colorPH}. We can see the convex
intruder associated with the region of negative specific heat on the caloric
curve, the minimum energy $E'_{c}$ at which the curve $S(E)$ presents a spike
(because $\delta E=0$ implies $\delta S=0$ since $\delta S=\beta_{\infty}\delta
E$),
and the branch of unstable states with lower entropy than the stable states
with the same energy. In the CE, the evolution
is more interesting. As we reduce the
temperature, starting from the gaseous phase, the system first undergoes a
phase transition at $T_c$ towards the condensed phase (fermion ball), followed
by a catastrophic collapse at $T'_c$ towards a black hole. The series of
equilibria is stable until $T_{c}$, becomes unstable between $T_c$
and $T_*$, is stable again between $T_*$ and $T'_c$ and becomes unstable
again afterwards. The free energy versus temperature curve is plotted in
Fig. \ref{Feta_R50_N0p399PH}. We can see the signal of a first order phase
transition at $T_t$ marked by the discontinuity of the first derivative of the
free energy (this first order phase transition does not take place in
practice as we have explained) and the two spinodal points $T_c$ and $T_*$ at
which the curve $F(T_{\infty})$
presents a spike (because $\delta  T_{\infty}=0$ implies $\delta F=0$ since
$\delta (\beta F)=E\delta \beta_{\infty}$) and where zeroth
order phase transitions (associated with the
discontinuity of the free energy) take place. We also see the minimum
temperature $T'_c$ below which the system collapses towards a black hole (the
curve
$F(T_{\infty})$ also presents a spike there). For large
values of $N$, the
caloric curve approaches the classical caloric curve of
Fig. \ref{kcal_N01_linked_colorsPH}.

{\it Remark --} The equilibrium states in the region of
negative specific heat have a core-halo structure with a quantum core (fermion
ball) surrounded by an isothermal atmosphere. They are unstable in the CE
(saddle points of free energy) and represent a ``critical droplet'' or a free
energy ``barrier'' that the system must cross in order to trigger a phase
transition from the gaseous phase to the condensed phase (a rare event as we
have seen). By contrast, these core-halo solutions are stable in the MCE
(maxima of entropy at fixed energy). They may describe DM halos with a quantum
core (representing a large bulge) and an isothermal atmosphere as suggested in
\cite{modelDM,mcmh}.

\subsection{$Z$-shape structure}

We now consider a radius in the range $R>R_{\rm MCP}=92.0$ so that the
caloric curve may display
both canonical and microcanonical phase transitions \cite{acf}. The description
of the CE
is the same as above so we focus on the MCE. 

For $N\ll N_{\rm OV}$ we are in the nonrelativistic
limit considered in \cite{ijmpb}. When $N<N_{\rm MCP}(R)$, where $N_{\rm
MCP}(R)\simeq 2.20\times 10^6/R^3$  corresponds
to a microcanonical
critical point, the
situation is the same as above.  When 
$N_{\rm MCP}(R)<N<N_{\rm OV}$ the caloric curve has a $Z$-shape structure
similar to Fig. 15 of \cite{ijmpb} (dinosaur's neck). There is a phase
transition from the gaseous
phase to the condensed phase. The condensed phase corresponds to a fermion ball
(similar to a white dwarf or a neutron star) surrounded by a hot and massive
atmosphere.\footnote{In the condensed phase the system has a
core-halo structure. The core is condensed and behaves as a completely
degenerate Fermi gas at $T=0$ (because $k_B T\ll\mu(r)$). For
physically relevant parameters we find that the condensed core contains about
$1/4$ of the total mass \cite{acf}. It has a very negative
potential energy. Since the total energy is conserved the halo must have a very
high kinetic energy, i.e., a very high temperature.}
As before, the physical phase
transition does not take place at the transition energy $E_t$ where the two
phases have the same
entropy but rather at $E_c$ (spinodal point), the energy
at which the metastable gaseous branch disappears and the system collapses
(gravothermal catastrophe). Once in
the condensed phase, if the energy increases, there is an explosion, reverse to
the collapse,
from the condensed
phase to the gaseous phase at $E_*$ (spinodal point). We can in this manner
generate an hysteresis cycle in the MCE \cite{ijmpb}. 

\begin{figure}
\begin{center}
\includegraphics[clip,scale=0.3]
{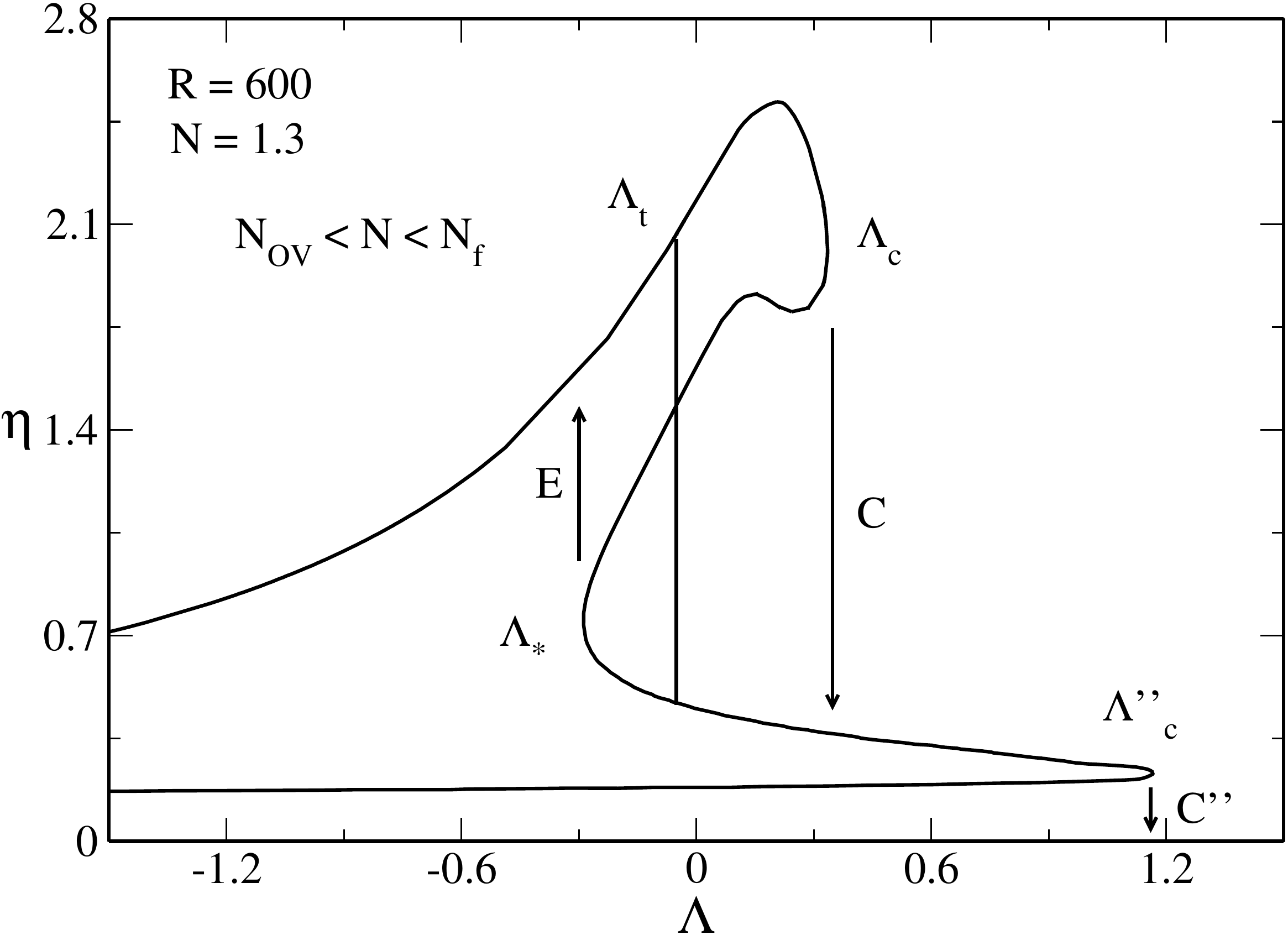}
\caption{Caloric curve for $N_{\rm OV}<N\ll N_{\rm max}$ (specifically
$R=600$ and $N
= 1.3$).  The arrows
refer to the MCE: (C) collapse at $E_c$ from the gaseous phase to
the condensed phase (fermion ball); (C'') collapse at $E''_c$ from the
condensed phase (fermion ball) to a black hole; (E) explosion at $E_*$ from the
condensed phase (fermion ball)  to the gaseous phase.}
\label{kcal_R600_N1p3_unifiedPH}
\end{center}
\end{figure}

\begin{figure}
\begin{center}
\includegraphics[clip,scale=0.3]{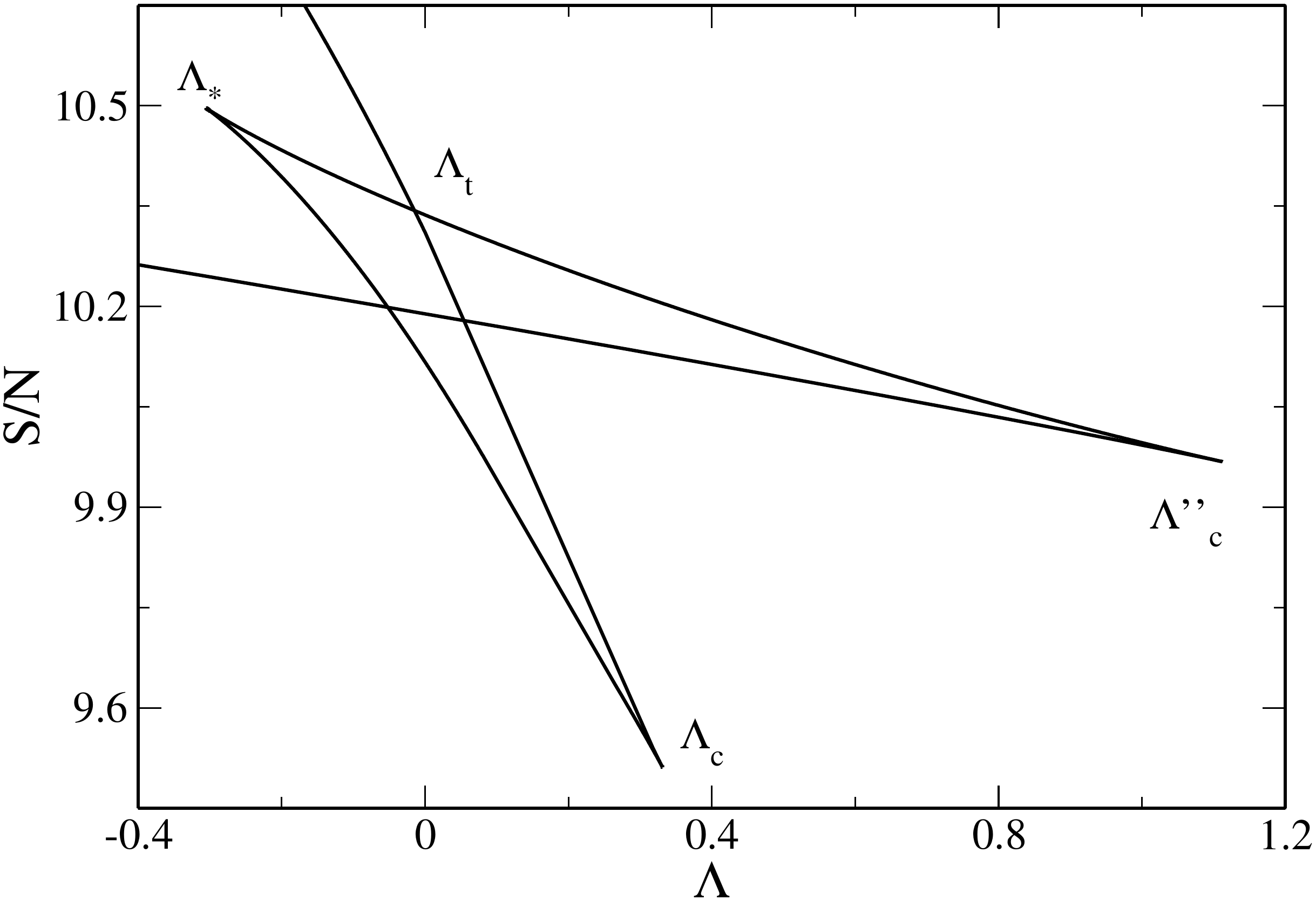}
\caption{Entropy per fermion as a function of the normalized energy for
$N_{\rm OV}<N\ll N_{\rm max}$ (specifically $R = 600$ and $N = 1.3$).}
\label{XSLambda_R600_N1p3PH}
\end{center}
\end{figure}

For  $N_{\rm
OV}<N\ll N_{\rm max}$, the caloric
curve
has the form of Fig. \ref{kcal_R600_N1p3_unifiedPH}.
As we reduce the energy, starting from the gaseous phase, the system first
undergoes a
phase transition at $E_c$ towards the condensed phase (fermion ball), followed
by a catastrophic collapse at $E''_c$ towards a black hole. The series of
equilibria is stable until $E_{c}$, becomes unstable between $E_c$
and $E_*$, is stable again between $E_*$ and $E''_c$ and becomes unstable
again afterwards. The entropy versus energy curve is plotted in
Fig. \ref{XSLambda_R600_N1p3PH}. We can see the signal of a first order phase
transition at $E_t$ marked by the discontinuity of the first derivative of the
entropy (this first order phase transition does not take place in
practice for the reason explained above) and the two spinodal points
$E_c$ and $E_*$, at which the curve $S(E)$
presents a spike and where zeroth order phase transitions (associated with the
discontinuity of the entropy) take place. We also see the minimum
energy $E''_c$ below which the system collapses towards a black hole (the curve 
$S(E)$
also presents a spike there). For large
values of $N$, the
caloric curve approaches the classical caloric curve of
Fig. \ref{kcal_N01_linked_colorsPH}.

\section{Astrophysical applications}
\label{sec_aa}

We can use our model to draw a physical scenario of stellar evolution
(of course in a very simplified setting) complementing the
results of \cite{pomeau1,pomeau2}. Let us consider
a gaseous star initially with a high energy and a high temperature. We consider
two types of
evolution, a canonical one correponding to Fig.
\ref {kcal_R50_N0p399_unifiedPH} and a microcanonical
one
corresponding to Fig. \ref{kcal_R600_N1p3_unifiedPH}. As the star radiates
light into space, its
temperature $T(t)$ (resp. energy $E(t)$) progressively decreases. As a result,
the star
follows the series of
equilibria on the gaseous branch until it reaches the minimum temperature
$T_c$ (resp. minimum energy $E_c$)
at which point it becomes unstable and collapses. The collapse continues until
quantum
mechanics comes into play. Then,  if its mass is not
too high, the star settles on a quiescent equilibrium
state,
a compact object similar to a white dwarf or a neutron star. The instability
at the critical point corresponds to
a saddle-center bifurcation. The evolution
of the radius $R(t)$ of the star on the gaseous branch can be described by a
Painlev\'e I equation of the form $\ddot R=-ct-bR^2$ which explains the
slow-fast transition in the life and death of a star \cite{pomeau1,pomeau2}. The
following discussion depends whether we work in the CE or in the MCE.

In the CE it can be shown that the
perturbation $\delta\rho(r)$ that triggers the instability at the critical
temperature $T_c$ has a core structure (one node) \cite{aaiso,pomeau1}
while the velocity perturbation $\delta v(r)$ has an implosive structure (no
node)
with $\delta v<0$ (implosion) \cite{pomeau1}. Therefore, the star is
expected to collapse as a whole  and
form a compact object (white dwarf or neutron star) containing all the mass
\cite{ijmpb,pomeau1,acf}. 
This phase transition is reminiscent of the hypernova phenomenon for
supermassive stars (above $40\, M_{\odot}$) which shows very intense and
directive gamma ray bursts,
but no
explosion of matter (or a very faint one).
When
$N$
is large enough ($N\gtrsim N_{\rm OV}$) there is no equilibrium state anymore,
and the
star collapses towards a black hole. 

In the MCE it can be shown that the
perturbation $\delta\rho(r)$ that triggers the instability at the critical
energy $E_c$ has a core-halo structure (two nodes) \cite{paddy,pomeau2}
while the velocity perturbation $\delta v(r)$ has an implosive-explosive
structure (one node)
with $\delta v<0$ in the core (implosion) and $\delta v>0$ in the halo
(explosion) \cite{pomeau2}. Therefore, the star is expected to split into a
collapsing core, leading ultimately to a compact object (white dwarf or neutron
star) containg a finite fraction ($\sim 1/4$) of the total initial mass
\cite{acf}, and
an explosive (hot) halo expanding  at large distances \cite{ijmpb,pomeau2,acf}.
This
phase
transition is
reminiscent of the red giant structure of stars with low or intermediate mass
(roughly $0.3-8\, M_{\odot}$) in a late phase of stellar evolution before the
white dwarf stage. The
implosion of the core and the explosion of the halo is also similar to the
supernova explosion of massive stars with mass in the range of $8-40\,
M_{\odot}$ resulting in the formation of a neutron star.  When $N$
is large enough ($N\gtrsim 4N_{\rm OV}$) there is no equilibrium state anymore,
and the
core collapses towards a black hole.

Similar results apply to stellar systems (globular clusters, galactic
nuclei,...) and dark matter
halos with, however, some differences. For these systems, only the
MCE makes sense. At the critical energy $E_c$, they undergo
the  gravothermal catastrophe \cite{lbw} and separate into a collapsing core and
a halo. 
They are in hydrostatic equilibrium but their evolution is
induced by the
temperature gradient between the core and the halo and by the fact that the
core has a negative specific heat $C_c<0$ \cite{lbw,thirring}. Therefore, by
losing heat, the core grows hotter, contracts,
loses heat again to the profit of the halo,  and evolves away
from equilibrium in an unstoppable process (thermal runaway). In the
case of classical objects like globular clusters, core collapse leads to a
binary star surrounded by a hot halo \cite{inagakilb}. In the case of
fermionic dark matter
halos where quantum mechanics becomes important at small scales, core collapse
may lead to a ``fermion ball'' (representing a quantum core
or a  large bulge) $+$ the expulsion of a hot
envelope. The gravitational energy released by the collapse of the
core heats up the envelope. In the box
model, the
atmosphere  is held by the walls of the box. Without the box, the atmosphere
is expelled at large
distances (see Figs. 38 and 41 of \cite{clm2} for an illustration in the
context of the fermionic King model). Therefore, in the MCE
the remnant is a
fermion
ball. Alternatively, in the case of galactic nuclei (star
clusters) \cite{zp,fit,strevue} and self-interacting dark matter
\cite{balberg}
the collapsing core during the gravothermal catastrophe may become relativistic
and finally experience a dynamical instability of general relativistic origin
leading to
a supermassive black hole of the right size ($10^6\le M/M_{\odot}\le 10^9$) to
explain quasars and active galactic nuclei (AGNs). The halo is
not crucially affected by the collapse of the core and maintains its initial
(isothermal) structure. The timescale
governing the phase transition of these different systems (stars and stellar
systems) is very different.
For supernovae where the energy is carried quickly by neutrinos they are fast,
a few days, but for globular clusters, galactic nuclei, and dark matter
halos they are very slow
(secular), of the order of the age of the Universe.

In summary, we suggest that the microcanonial phase transition
occurring in the
self-gravitating Fermi gas  may be related to the onset of red giant
structure
or to the supernova phenomenon. In these spectacular events, the collapse of the
core of the system (resulting ultimately in the formation of a white dwarf or a
neutron star) is accompanied by the explosion and the expulsion of a hot
envelope. Newtonian gravity is sufficient to describe the white
dwarf and the planetary nebula that follow the red giant stage while general
relativity is necessary to describe neutron
stars or black holes formed from supernova explosion.  Similarly, dark matter
halos made of fermions may have a core-halo structure where the core may be a
fermion ball (bulge) or a black hole depending on the mechanism at
work.

{\it Remark --} It is
interesting to compare our results with previous results obtained in the same
context. Lynden-Bell and Wood
\cite{lbw}, considering a classical self-gravitating gas in the MCE,
found the emergence of a core-halo structure and related it to the
onset of red giants. Hertel and
Thirring \cite{ht}  argued that, through the
electromagnetic radiation, a star is
in heat contact with the rest of the Universe which acts as heat bath. As
a result they worked in the CE and mentioned
the analogy between the canonical  phase transition and the formation
of supernovae. However, this analogy may not be fully correct because the
phase
transition that they obtained just corresponds to an implosion. This is because
they worked in the CE and considered relatively small systems
while
the phase transition leading to an implosion-explosion
phenomenon, associated with a core-halo structure, occurs in the MCE
for larger systems. Chandrasekhar, in the last sentence of \cite{sch},
suggested that the supernova
phenomenon may result from the inability of a star of mass greater than  $M_{\rm
max}$ to settle down to the final state of complete degeneracy without
getting rid of the excess mass. He assumed that sufficient mass is ejected
so that the remnant has a mass $M_{\rm core}<M_{\rm max}$. He did not
anticipate the formation of a black hole when $M_{\rm core}>M_{\rm max}$.
At a qualitative level, our scenario encompasses and generalizes these original
ideas.

\section{Conclusion}

In this Letter, we have presented the main results of our recent
investigations concerning the statistical mechanics of self-gravitating fermions
at finite temperature in the framework of general relativity. Additional results
and details can be found in our companion papers \cite{acb,acf,rgf,rgb}.
In particular, we have obtained in Ref.  \cite{acf}  the complete
phase diagram of the general relativistic Fermi gas in the $(N,R)$
plane determining the canonical and microcanonical critical points of this
system,
thereby generalizing the Newtonian results of \cite{ijmpb}. This
phase diagram was useful, for example, to interprete the results of
Roupas and Chavanis \cite{rc} who studied phase transitions in the general
relativistic Fermi gas using different control parameters, $GNm/Rc^2$
and $R/R_{\rm OV}$, instead of
$N/N_{\rm OV}$ and $R/R_{\rm OV}$. 

We have also discussed astrophysical applications of our thermodynamical
approach. In particular, we have argued that the CE may be
relevant to describe
the life and death of supermassive stars which collapse (implode) without
exploding (hypernova phenomenon) while the MCE
may be relevant to describe the life and death of less massive stars which
present a more complex evolution marked by the  collapse  (implosion) of the
core and the explosion of the halo (supernova phenomenon)
\cite{pomeau1,pomeau2}. This is an interesting consequence of ensembles
inequivalence for systems with long-range interactions \cite{campabook}. Other
interesting applications of the fermionic model in relation to dark matter halos
made of massive neutrinos are given in \cite{btv,clm2,vss,rar}.

\end{document}